\documentclass[9pt,twocolumn,twoside]{osajnl}

\journal{ol} 
\usepackage{setspace,gensymb,placeins,afterpage,array,graphicx}
\newcolumntype{P}[1]{>{\centering\arraybackslash}p{#1}}
\setboolean{shortarticle}{false} 
\ifthenelse{\boolean{shortarticle}}{\colorlet{color2}{color2b}}{\colorlet{color2}{color2}} 

\title{All-optical tuning of a diamond micro-disk resonator on silicon} 

\author[1,2]{Paul Hill}
\author[3]{{Charalambos Klitis}}
\author[1]{Benoit Guilhabert}
\author[3]{Marc Sorel}
\author[1]{Erdan Gu}
\author[1]{Martin D. Dawson}
\author[1]{Michael J. Strain,*}

\affil[1]{Institute of Photonics, Dept. of Physics, 99 George St., Technology and Innovation Centre, University of Strathclyde, Glasgow, G1 1RD}
\affil[2]{Diamond Science and Technology, Centre for Doctoral Training, University of Warwick, Gibbet Hill Road, Coventry, CV4 7AL}
\affil[3]{School of Engineering, University of Glasgow, Glasgow G12 8LT}

\affil[*]{Corresponding author:michael.strain@strath.ac.uk}

\dates{Compiled \today}

\doi{\url{ }}

\begin{abstract}
High quality integrated diamond photonic devices have previously been demonstrated in applications from non-linear photonics to on-chip quantum optics.  However, the small sample sizes of single crystal material available, and the difficulty in tuning its optical properties, are barriers to the scaling of these technologies.  Both of these issues can be addressed by integrating micron scale diamond devices onto host photonic integrated circuits using a highly accurate micro-assembly method.  In this work a diamond micro-disk resonator is integrated with a standard single mode silicon-on-insulator waveguide, exhibiting an average loaded Q-factor of 3.1$\times$10$^{4}$ across a range of spatial modes, with a maximum loaded Q-factor of 1.05$\times$10$^{5}$.  The micron scale device size and high thermal impedance of the silica interface layer allow for significant thermal loading and continuous resonant wavelength tuning across a 450 pm range using a mW level optical pump. This diamond-on-demand integration technique paves the way for tunable devices coupled across large scale photonic circuits.
\end{abstract}
\setboolean{displaycopyright}{false}
\begin{document}
\maketitle
\thispagestyle{fancy}
\ifthenelse{\boolean{shortarticle}}{\ifthenelse{\boolean{singlecolumn}}{\abscontentformatted}{\abscontent}}{}
\section{Introduction}
Diamond, in single crystalline form and with its large palette of potential colour centres, is a particularly attractive optical material for applications ranging from high resolution magnetometry \cite{Casola2018} to quantum information processing \cite{Hanson2014,10qubit}. The use of integrated photonics to strongly confine optical fields in single crystal diamond (SCD) has underpinned a wide range of key demonstrations where strong light-matter interaction is crucial, including Raman lasing \cite{Latawiec2015}, Purcell enhancement of single photon emitters \cite{Englund2015}, optomechanics \cite{LoncarMechanics} and non-linear optics \cite{Hausmann2014d}.  Typically these devices are fabricated by integrating a piece of SCD with a secondary material for structural support and to allow definition of waveguides in the diamond itself \cite{Hausmann2012e, Piracha2016a}, or through evanescent field interaction to guided wave structures in the complementary material \cite{Thomas2014d}.  In both cases the total footprint of the diamond photonic circuit is limited by the extent of the available SCD material, typically in the order of mm$^{2}$.  The commonly wedged thickness of SCD chips \cite{Gao2016} can also be a barrier to scaling, with geometry variations across a single chip affecting optical performance of nominally identical devices, preventing the design of integrated circuits without complex pre-compensation.  Crucially, the ability to actively tune integrated photonic device performance is extremely limited in diamond as it presents high thermal conductivity ($\approx$2000 W/m.K) \cite{Kaminskii_2006}, a low thermal coefficient of refractive index (~1.5 $\times$10$^{-5}$)  \cite{Mildren_2013}, and no significant second order non-linearity.  Current tunable optical devices use demanding methods such as mechanical deformation of nano-beams \cite{Loncar2019} or environmentally induced refractive index modulation \cite{Zhang2018}.  

\begin{figure}
	\centering
	\includegraphics[width=0.45\textwidth]{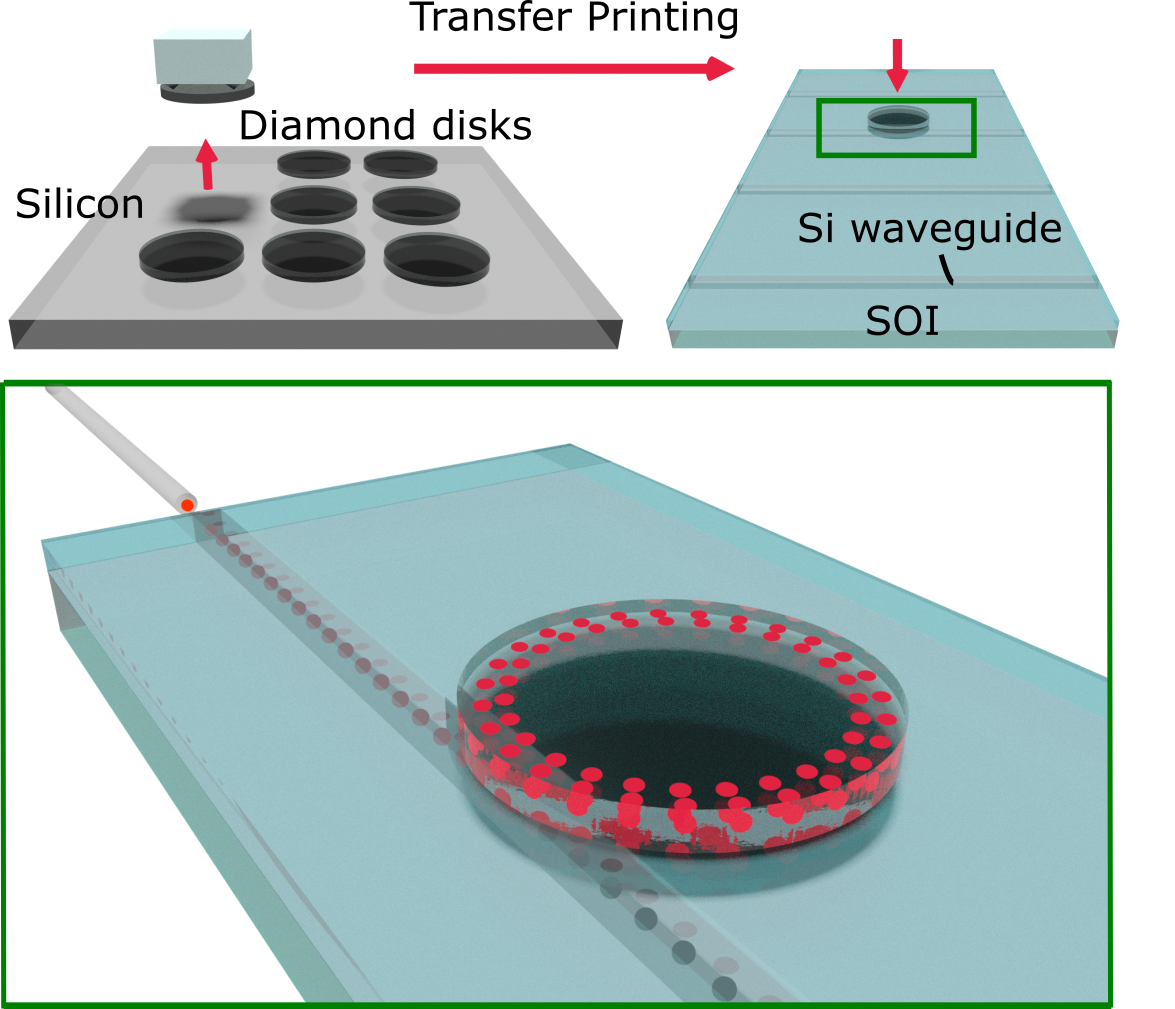}
	\caption{Schematic of a hybrid integration scheme where diamond micro-disk resonators are fabricated separately from a host Photonic Integrated Circuit chip.  The fully fabricated diamond resonators are transferred onto the silicon photonic chip using a high accuracy transfer printing method.}
	\label{Schematic}
\end{figure}
In this work we present a method for the integration of high quality SCD devices with pre-fabricated Photonic Integrated Circuits (PICs) on a second material platform, based on micro-assembly. By creating transferable, monolithic diamond devices, the limitation of the SCD substrate size is lifted, allowing a diamond on-demand hybrid optical system design.  Furthermore,  the diamond micro-resonators presented here are directly printed onto silica using an adhesiveless process.  This produces a high thermal resistance interface between the diamond and its host silicon substrate, allowing for relatively high local temperatures to be supported in the diamond material.  It is found that direct optical pumping of lossy resonant modes is sufficient to tune the material refractive index through the thermo-optic effect using mW level pump powers.  A schematic of the integration scheme is shown in Figure \ref{Schematic}, where diamond devices are pre-fabricated on a donor substrate before integration with a host PIC chip using an accurate transfer printing technique \cite{McPhillimy2018,Guilhabert2018}. 

 \section{Methods}
\subsection{Diamond membrane fabrication and printing}
\begin{figure}
\centering
\includegraphics[width=\linewidth]{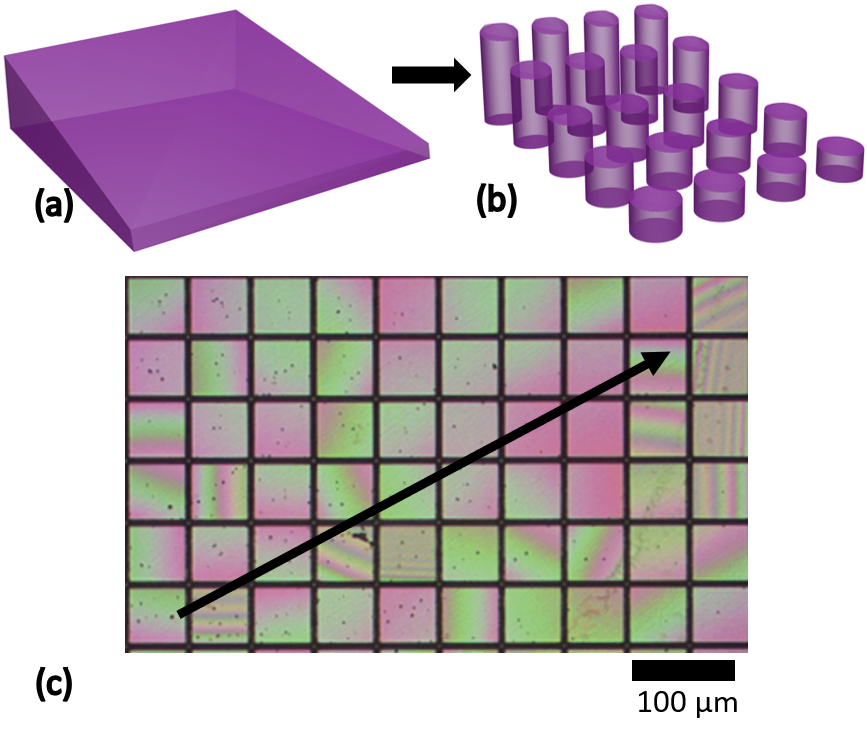}
\caption{a) A schematic of a wedged diamond sample, b) patterned devices illustrating device thickness selection using an iterative print and thin process.  c) Optical micrograph of an array of tessellated squares fabricated on a wedged single crystal diamond sample.  The arrow shows the major wedge axis.}
\label{WedgeDisk}
\end{figure}

The hybrid integration technique presented in this work is based on a micro-transfer printing method where diamond devices and their host PICs are fabricated separately and assembled using an accurate pick and place tool \cite{McPhillimy2018}.  In this case the diamond is integrated with a silicon-on-insulator (SOI) chip, but it is equally applicable to other material systems.  In order to ensure high optical mode overlap with the diamond material, disk resonators are used to evanescently couple to the silicon bus waveguide, as shown schematically in Figure \ref{Schematic}.  The SCD samples are fabricated using laser dicing and polishing to obtain square pieces with 2 mm side-lengths and a thickness of around 30 $\mu$m.  The typical wedge gradient of these pieces is in the order of 2.75 $\mu$m$/$mm.  In order to isolate the effect of this wedge, monolithic diamond devices can be defined across the chip as shown schematically in Figure \ref{WedgeDisk}.  Devices of the desired thickness can then be selected from the array for integration.  This method allows usage of the whole array through an iterative process of printing target thickness devices, globally etching the full sample until the next set of devices are within tolerance and repeating the printing.  Fabrication of the thin film SCD samples was carried out following our previous work using a lithography and Reactive Ion Etching (RIE) process\cite{Hill2018}. The sample is initially thinned to a few microns using an Ar$/$Cl$_2$ etch. The sample is then transferred onto a silicon carrier chip for patterning. An etch mask is patterned in hydrogen silsequioxane (HSQ) resist using electron beam lithography to ensure smooth sidewall features. Transfer of the pattern into the diamond is achieved using a high platen power Ar$/$O$_{2}$ ICP etch\cite{Hill2018}. After patterning, the hard-mask can be removed using a CF${_4}$ + H$_{2}$ etch and the membrane globally thinned using the Ar$/$Cl$_2$ etch to the desired thickness, measured using optical profileometry.

\begin{figure}
\centering
\includegraphics[width=\linewidth]{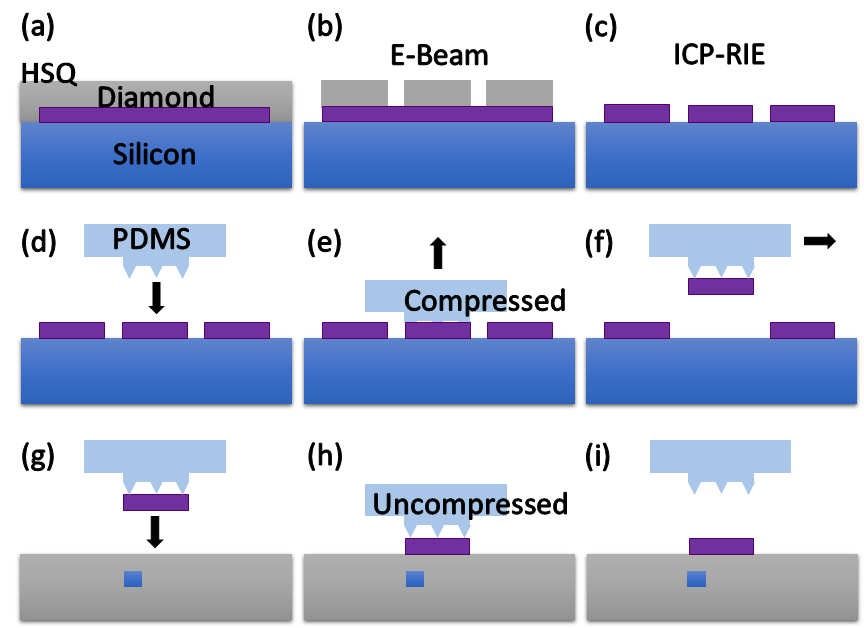}
\caption{Transfer printing process: a) resist is spun and b) patterned on a diamond membrane, c) the pattern is transferred to the diamond using inductively-coupled-plasma reactive ion etching. d)-f) A PDMS stamp is aligned with the diamond chiplet, brought into close contact and retracted to release the chiplet. g)-i) The chiplet is aligned over a host substrate, brought into contact and released, leaving the it transferred on the new substrate.}
\label{TP}
\end{figure}

A sample was patterned with an array of 49$\times$49 $\mu$m tessellated squares, separated by 1$\mu$m gaps to assess the printing process.  The major wedge axis lay along the diagonal of the sample as shown in Figure \ref{WedgeDisk}(c).
Once the free standing diamond devices are fully etched they can be detached from the silicon carrier substrate using a soft polymer stamp and transferred onto the host substrate. A schematic of the printing process is shown in Figure \ref{TP}. Firstly a Polydimethylsiloxane (PDMS) stamp is brought into contact with a target chiplet, e.g. one square of the tessellated array, and then retracted with a velocity above the critical velocity for object pickup \cite{Rogers2012}.  The chiplet adheres to the surface of the stamp and is released from the donor substrate.  The object is then positioned over the host substrate and brought into contact.  The stamp is removed at a velocity below the critical value, leaving the chiplet adhering to the surface. In this work the smooth surfaces of the diamond chiplets and host substrates, $<$nm r.m.s. roughness \cite{Gu2008}, means that printing can be achieved without an intermediate adhesion layer.

\begin{figure}
	\centering
	\includegraphics[width = \linewidth]{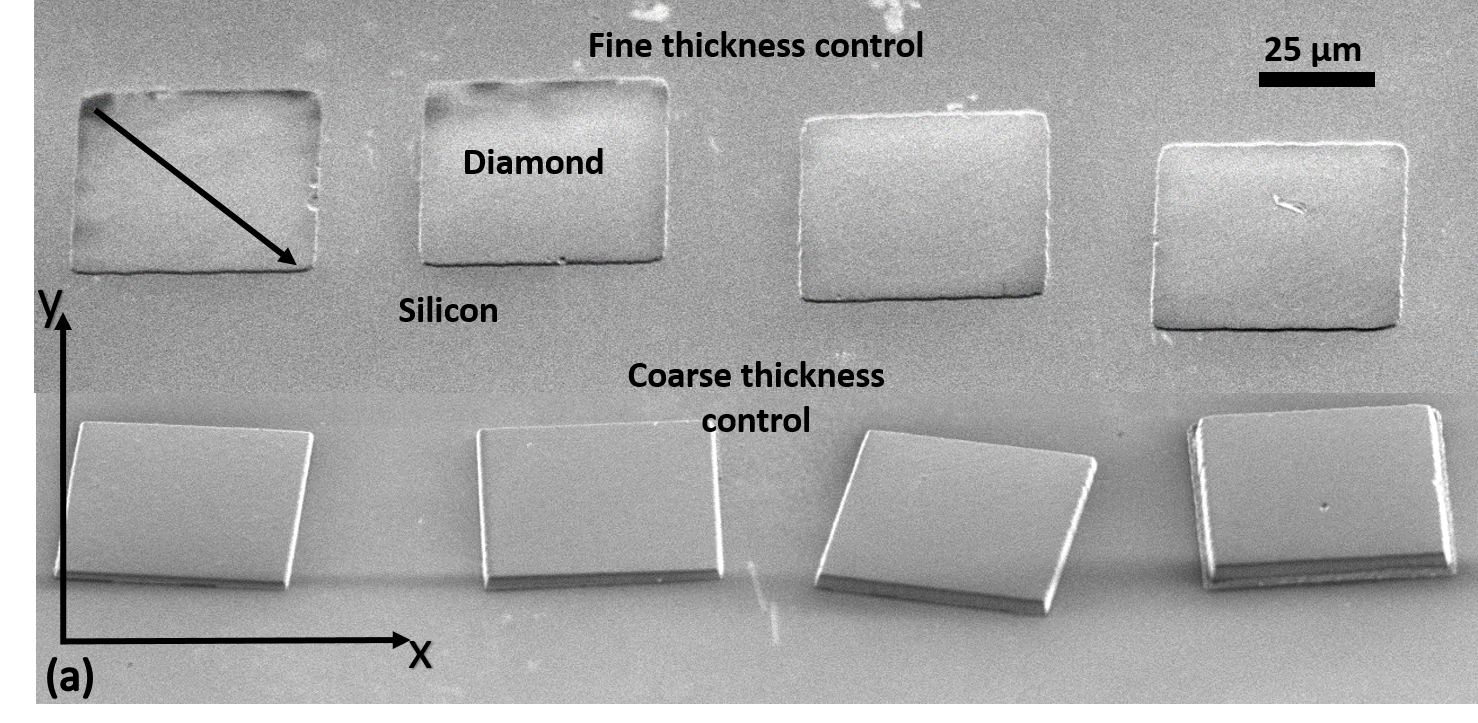}
	\includegraphics[width = 0.8\linewidth]{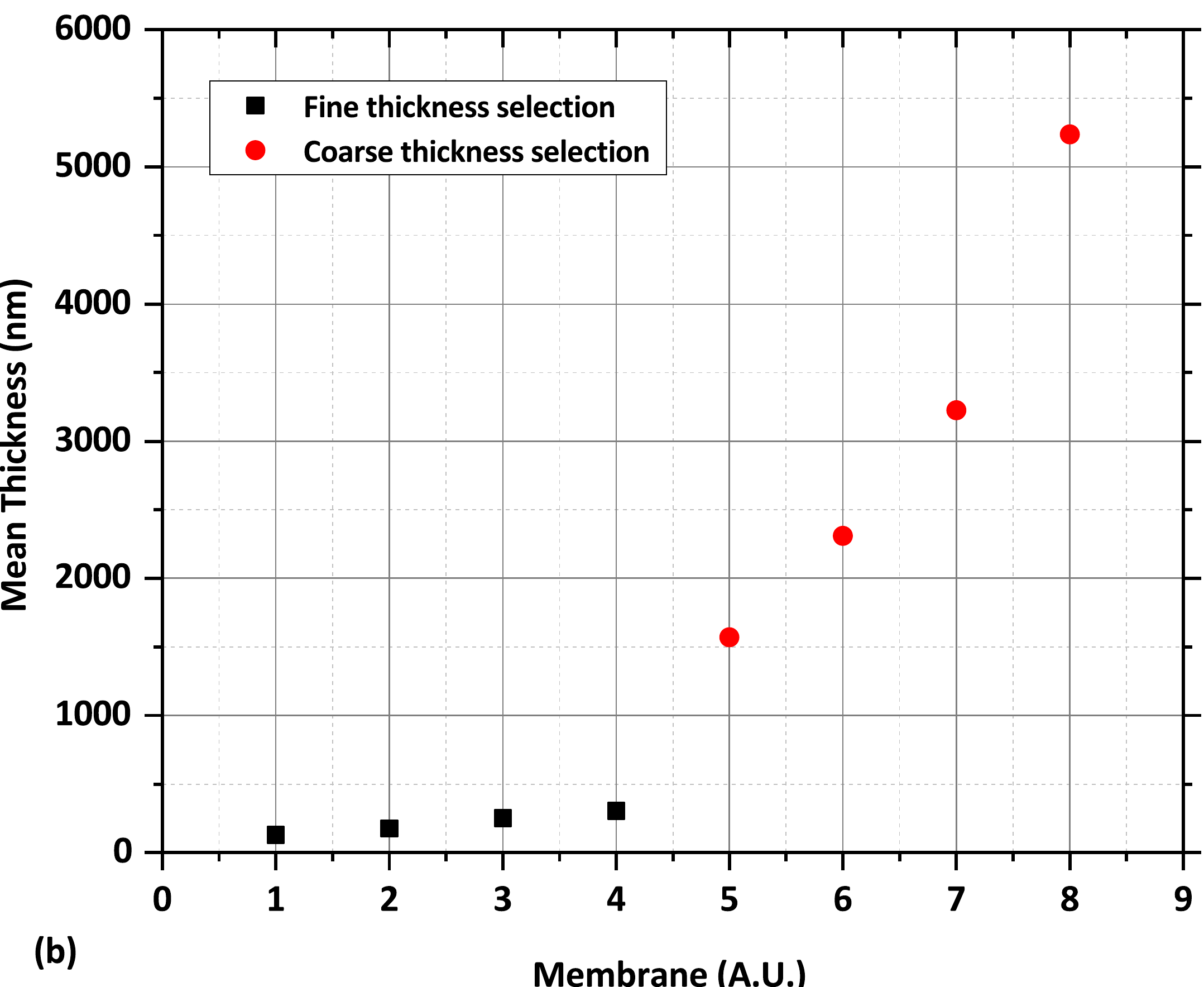}
	\caption{a) Two 45 deg tilted-view SEM micrographs combined to show 8 membranes that were printed for AFM measurements, the black arrow indicates the direction of largest thickness gradient. b) A plot showing the mean AFM heights for each of the 8 membranes demonstrating the ability of fine and coarse thickness selection.}
	\label{Thickness}
\end{figure}

Four chiplets were selected along both the maximum and minimum material wedge directions and printed onto a host piece of silicon. Their thickness was measured using atomic force microscopy (AFM). The difference of thickness between each chiplet's highest and lowest corners was used to determine a gradient over the diagonal length, which was found to match the global gradient of the diamond sample measured before pattering. SEM images of the printed chiplets are presented in Figure \ref{Thickness}(a).
Figure \ref{Thickness}(b) shows the measured average thickness of the eight printed chiplets, showing the ability to select for thickness across a wedged sample. The square with lowest average thickness was measured with a minimum corner height of $\approx$ 10 nm, demonstrating the suitability of the printing technique for delicate thin films.

\subsection{Diamond micro-disk integration}
A particular benefit to using transfer printing to assemble micro-photonic devices is that the geometry is not limited to what can be realised in single planar layer \cite{Rogers2012,McPhillimy2018}.  In this case a diamond micro-disk resonator was printed onto a silicon bus waveguide with an upper-cladding of silica, giving control over the coupling coefficient between the disk and waveguide using both the vertical and lateral separation of the devices.  The diamond disk was defined with a radius of 12.5 $\mu$m and a thickness of 1.8 $\mu$m, written into a HSQ electron beam resist. Following patterning, the disk was fabricated in the same way as the tesselated square array.  The silicon bus waveguide was fabricated on a 220 nm thick silicon-on-insulator material platform, with a width of 500nm.  An uppercladding of HSQ was spin coated onto the chip with a thickness of 250 nm.  The silicon waveguide was terminated with an inverse taper and embedded in a SU8 waveguide to allow off-chip coupling to fibre with low loss.

\begin{figure}
\centering
\includegraphics[width=0.85\linewidth]{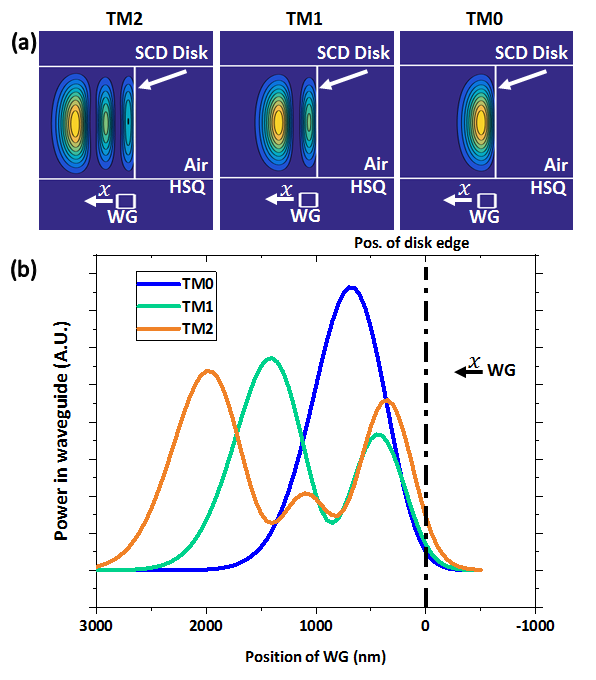}
\caption{a) Power distributions of the first three TM whispering gallery modes of a diamond disk resonator with 1.8 $\mu$m thickness printed on a silica substrate. b) A plot of the percentage of power present in the location of a silicon waveguide as a function of lateral offset, x, to the edge of the disk.}
\label{confinement}
\end{figure}

\begin{figure}
\centering
\includegraphics[width=0.7\linewidth]{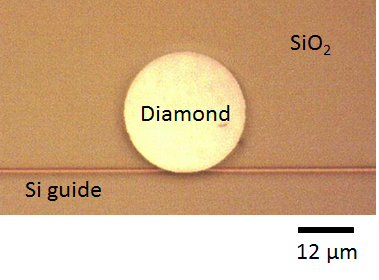}
\caption{An optical microscope image showing a 12.5 $\mu$m radius diamond disk integrated with a silicon waveguide using micro-transfer printing.}
\label{DiskPrint}
\end{figure}

The micro-disk dimensions allow multiple spatial modes in the cavity that will exhibit different propagation losses and coupling coefficients to the bus waveguide, which is a necessary component for the pump/probe optical cavity tuning presented here. Figure \ref{confinement} shows calculated mode profiles for the first three radial TM modes of the cavity calculated using a finite-difference-eigenmode solver. The parameter $x$ in Figure \ref{confinement} refers to the lateral offset between the edge of the disk resonator and the edge of the silicon waveguide it is being printed onto.  By varying this offset, the coupling coefficient between the waveguide and the disk can be controlled. Figure \ref{confinement}(b) shows the variation of the optical power confined to the waveguide area for these modes, showing modification of the ratio of coupling coefficients between them  can be achieved by variation of the lateral offset between the disk and waveguide.  In this work the diamond micro-disk was printed with its edge aligned to that of the silicon bus waveguide.  An optical microscope image of the printed diamond micro-disk on the silicon waveguide is shown in Figure \ref{DiskPrint}.

\subsection{Measurement setup}
\begin{figure}
	\centering
	\includegraphics[width = \linewidth]{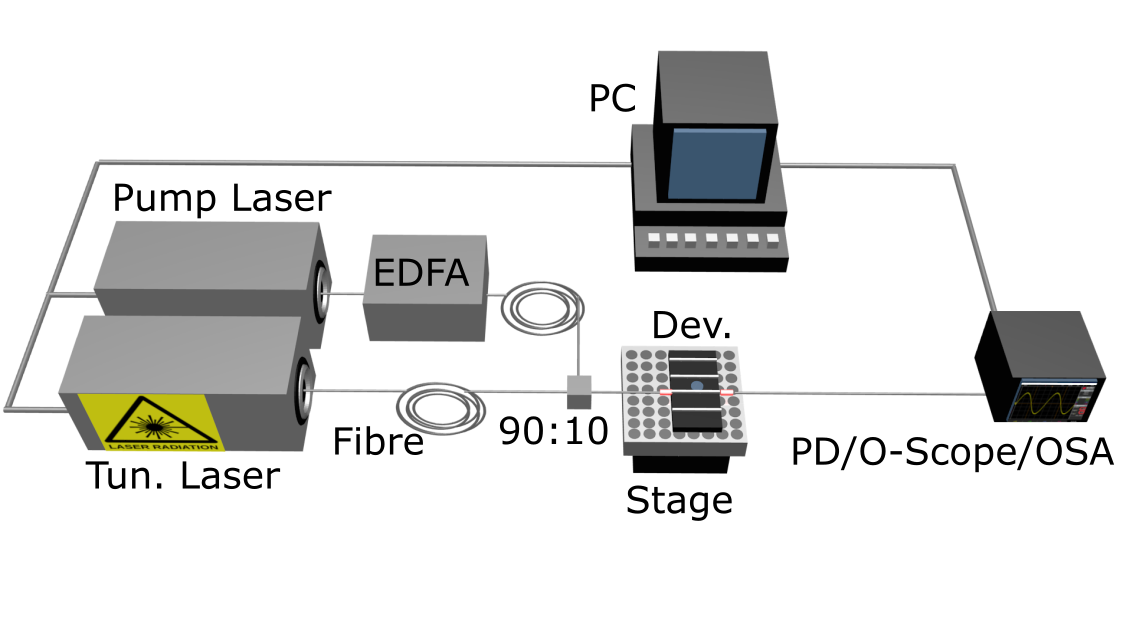}
	\caption{Optical measurement setup used for spectral characterisation of the integrated micro-disk.  The pump laser source, EDFA and OSA are onsly used for the pump/probe thermal tuning measurements.}
	\label{PumpSetup}
\end{figure}

The spectral characterisation of the micro-disk resonator and the optical tuning were both realised with the same measurement setup, as shown in Figure \ref{PumpSetup}.  For optical transmission measurements a tunable laser was coupled to the silicon chip using a lensed, polarisation-maintaining fibre. The output light was coupled to a second lensed fibre and collected with a photodiode and oscilloscope. 

For the thermo-optic tuning measurements, the setup was augmented with a second laser source that was amplified using an Erbium Doped Fibre Amplifier (EDFA) and multiplexed with the low power tunable laser source through a 90:10 fibre coupler to a lensed fibre. A second lensed fibre was used to collect the light from the output facet of the chip and coupled to an Optical Spectrum Analyser (OSA).  

\section{Results}
\subsection{Micro-disk transmission measurements}
\begin{figure}
	\centering
	\includegraphics[width = 0.9 \linewidth]{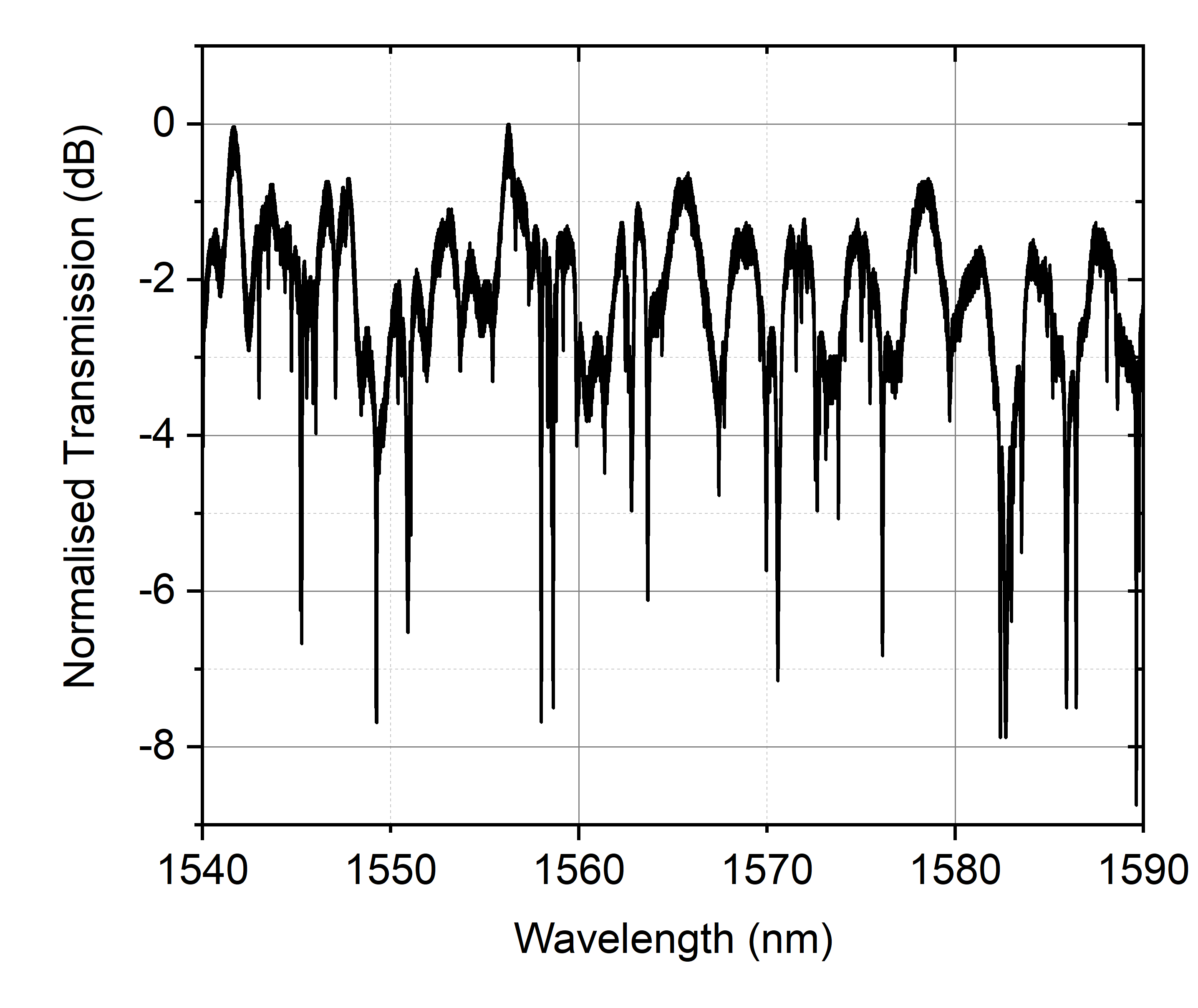}
	\caption{Normalised transmission spectrum of the hybrid diamond-silicon micro-disk resonator.}
	\label{transmission}
\end{figure}

\begin{figure}
\centering
\includegraphics[width = 0.8 \linewidth]{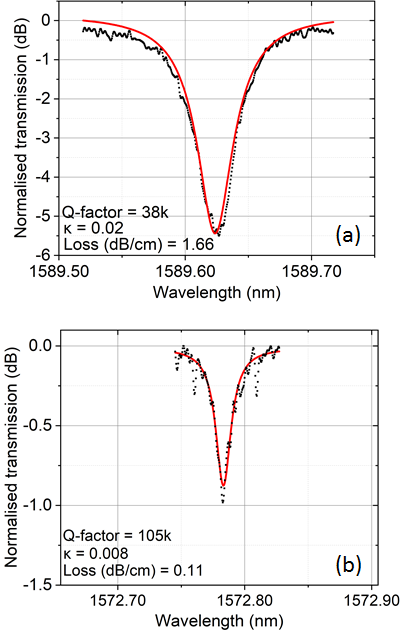}
\caption{a) Measured transmission and fit to analytic all-pass resonator function for a mode around the average loaded Q-factor of the device. b) Measured transmission and fitted curve for the highest measured loaded Q-factor resonance. $\kappa$ is the power cross-coupling coefficient and the loss refers to the distributed propagation loss value of the resonator.}
\label{resonance}
\end{figure}

\begin{figure}
\centering
\includegraphics[width = \linewidth]{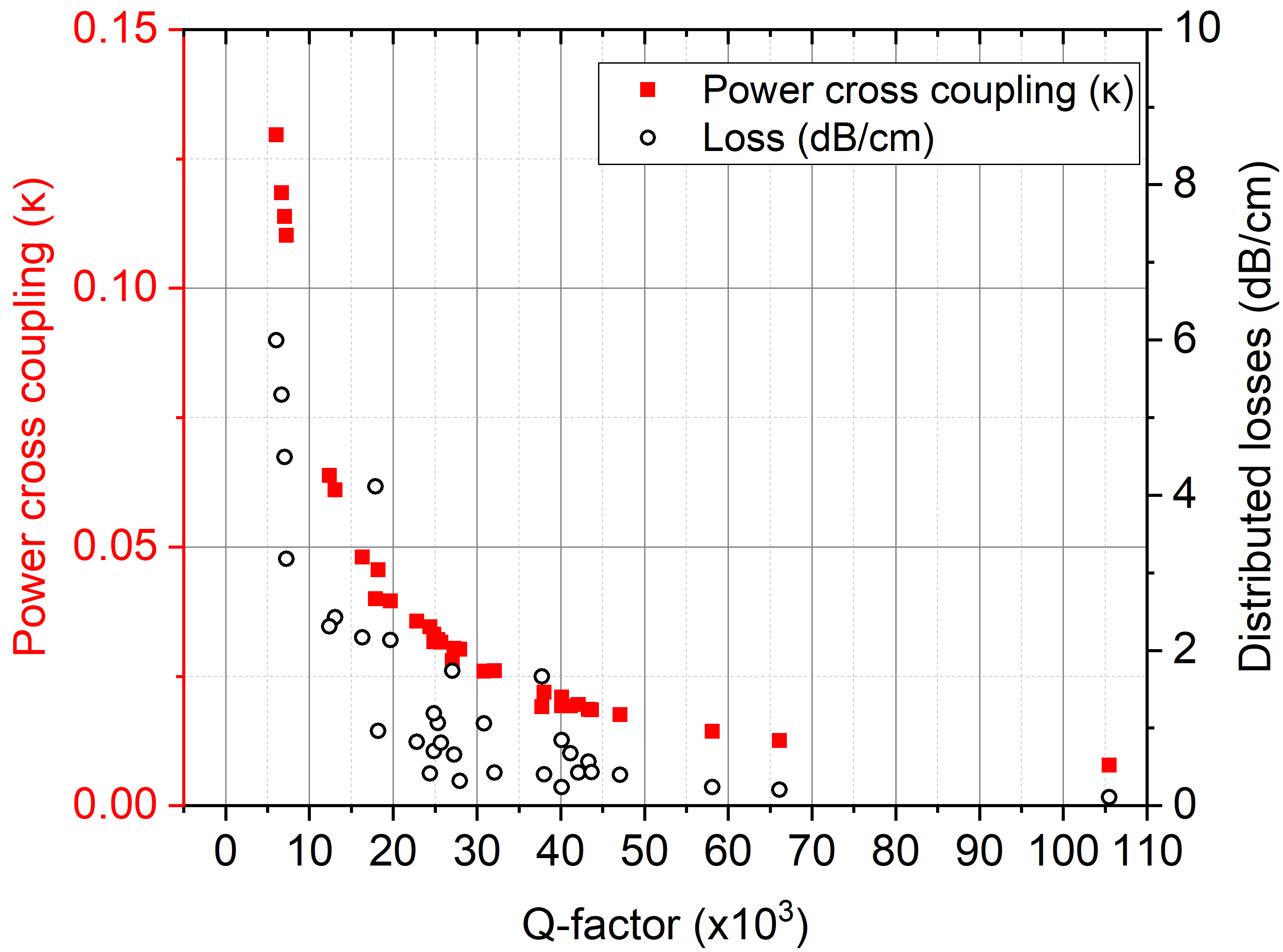}
\caption{Power cross-coupling coefficients (squares) and distributed losses (circles) as a function of measured loaded Q-factor.}
\label{Qfactors}
\end{figure}

A transmission spectrum of the diamond micro-disk resonator coupled to the silicon bus waveguide is shown in Figure \ref{transmission}.  As expected from the multiple spatial modes supported by the micro-disk geometry, the spectrum exhibits a large number of resonances.  
The large number of modes measured makes it difficult to identify a particular spatial mode solution.  Nevertheless, each resonance can be fitted to an analytic model for an all-pass, whispering gallery resonator to extract values for the coupling coefficient, round-trip propagation loss, loaded and intrinsic Q-factors \cite{Hill2018}.  An average loaded (intrinsic) Q-factor of > 3.1$\times$10$^{4}$ (1.90$\times$10$^{5}$) was found - with a representative example of a fitting plotted in Figure \ref{resonance}(a). There were a number of measured Q-factors significantly higher than the average, with the largest loaded (intrinsic) Q-factor resonance plotted in Figure \ref{resonance}(b) with a value of 1.05$\times$10$^{5}$ (9.96$\times$10$^{5}$). 
These values are comparable with other diamond resonator devices that have been reported in the literature\cite{Hausmann2012d,Latawiec2015}, showing that the transfer printing process does not induce significant additional optical loss in the resonators. All of the extracted power cross coupling coefficients and distributed losses across the spectrum are plotted against loaded Q-factor in Figure \ref{Qfactors}. As expected, the modes with higher Q-factors have the lowest losses and cross-coupling coefficients.  The correlation between propagation loss and coupling coefficient is likely due to the fact that modes where the overlap of the optical mode with the edge facet of the resonator is low, will experience less loss and will have a smaller overlap with the bus waveguide, reducing the coupling coefficient.  
\subsection{Thermo-optic tuning}
The micro-assembly of diamond micro-disks onto insulator provides a mechanism for optically tuning the device resonant wavelengths.  Absorption of light results in thermal energy being deposited in the waveguiding material.  The resultant increase in temperature induces a material refractive index shift based on the thermo-optic coefficient of index \cite{Lipson}. For devices fabricated in millimetre size diamond films, the large thermal conductivity ($\approx$2000 W/m.K) of the material and its low thermo-optic coefficient ($\approx$1.5 $\times$10$^{-5}$) mean that with typical on-chip power levels in the mW range, resonance tuning is extremely limited.  Resonators improve the tuning capability by locally trapping the optical mode and therefore producing higher local temperatures.  In the hybrid geometry presented here the diamond micro-disk is thermally isolated on the silica cladding of the host chip.  Therefore, any thermal energy deposited in the disk can only convect to the surrounding air, conduct through the substrate or radiate from the surface.  Convection and radiation are both low efficiency processes for diamond devices, the latter due to a small emissivity coefficient of the material.  The thermal conductivity of silica is 1.5 W/m.K, providing good thermal isolation of the diamond micro-disk.  Therefore, in this geometry, the combined effect of the optical mode confinement and the thermal isolation of the small diamond resonator, allows for significant temperature increases in the diamond with mW level optical pumping.

\begin{figure}
	\centering
	\includegraphics[width = 0.75\linewidth]{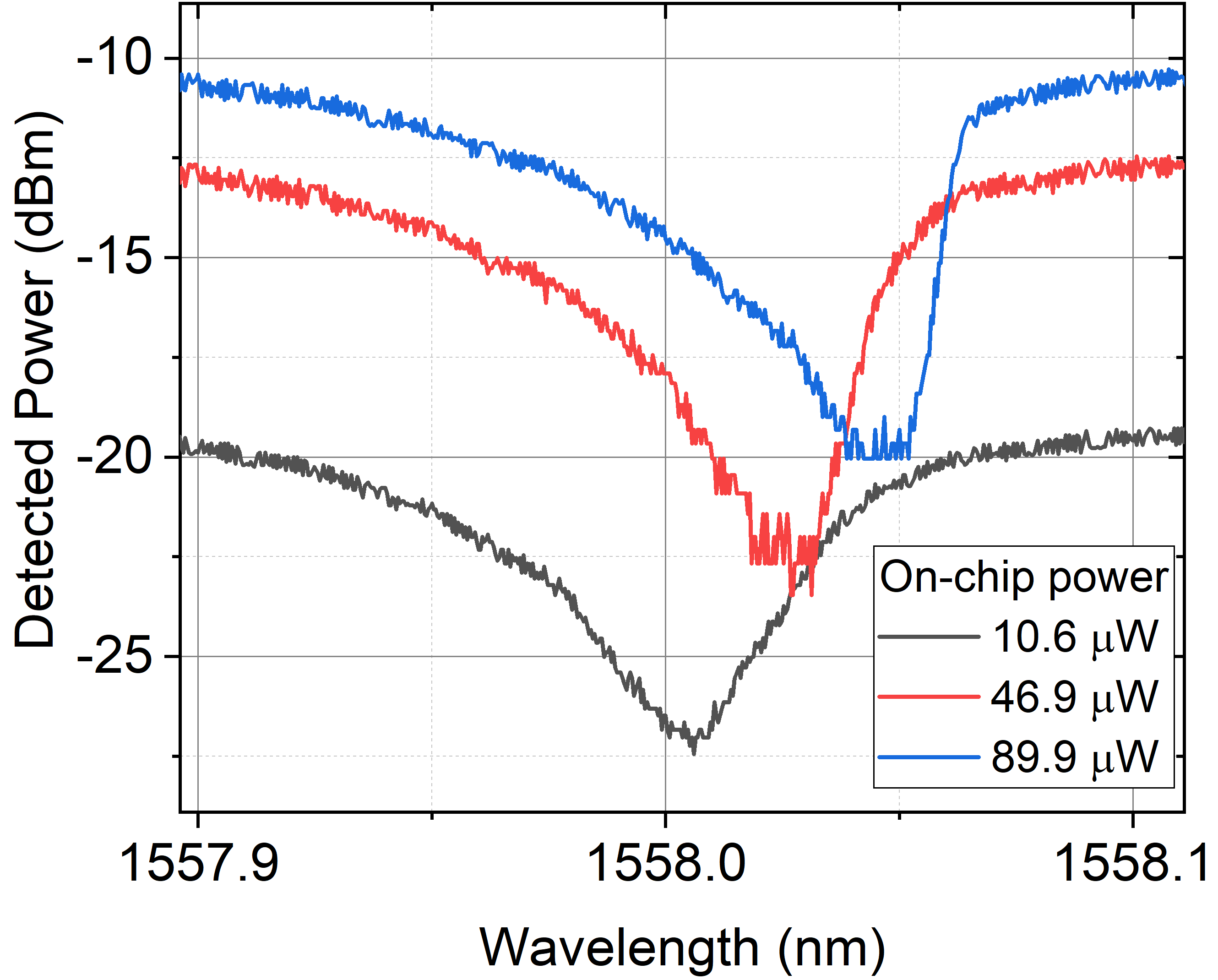}
	\caption{Transmission spectra of the diamond resonator measured using a continuously swept laser source with on-chip source power as a parameter.}
	\label{PowerDependence}
\end{figure}

The effect of increasing optical injection power on the resonator refractive index can be measured using the well known thermo-optic bistabilty in optical resonators \cite{Lipson}.  By sweeping the tunable laser source across a resonance from blue to red wavelengths, an asymmetric resonance response is recorded.  The transmission minimum can be used to calculate the peak resonance shift and therefore the thermally shifted refractive index and temperature of the device. Figure \ref{PowerDependence} shows three transmission spectra at different on-chip optical power levels, exhibiting a thermo-optic bistability.  

\begin{figure}
	\centering
	\includegraphics[width = 0.8 \linewidth]{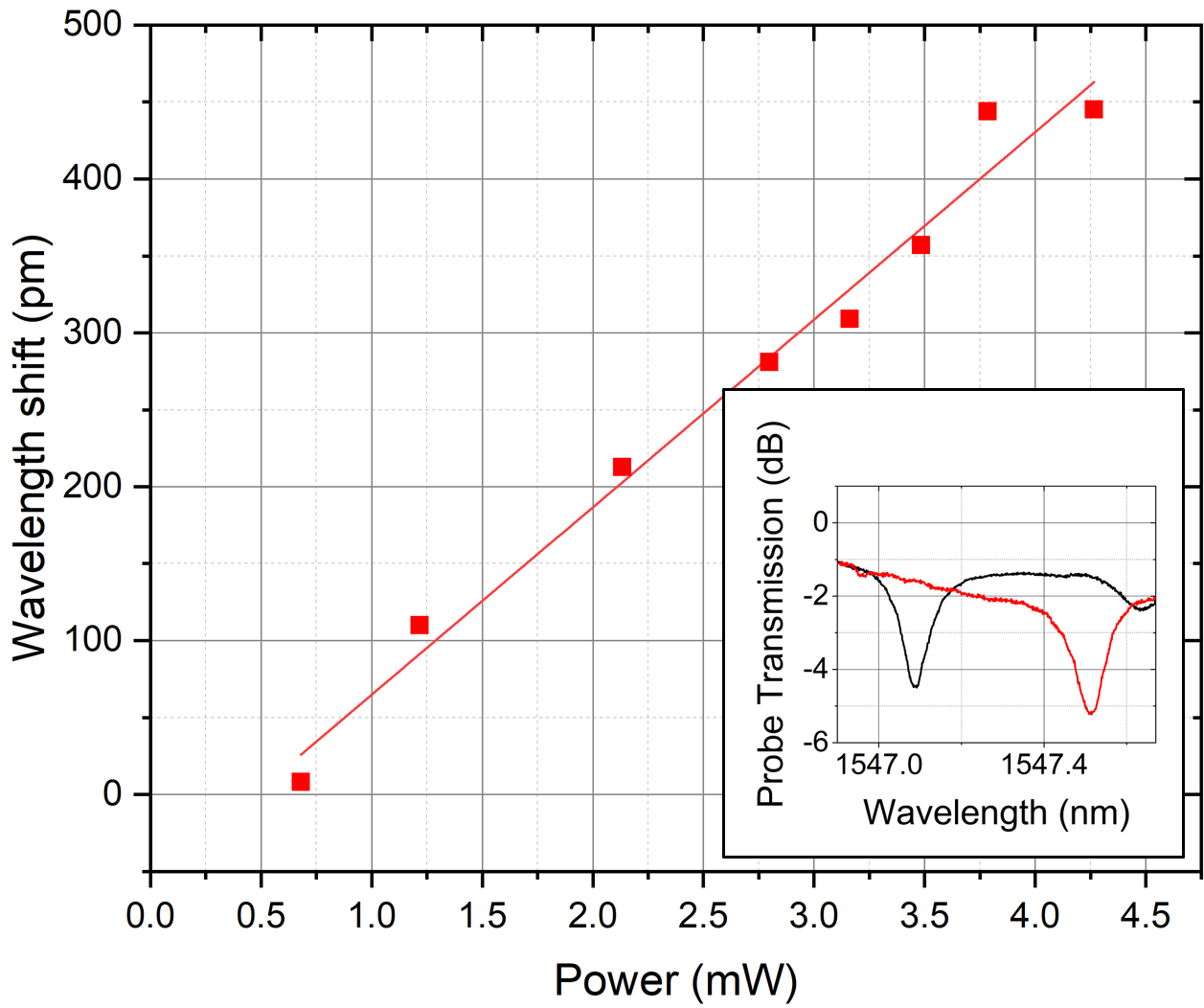}
	\caption{Probe measured wavelength shift of a resonance centred at \textasciitilde 1547 nm for different pump powers injected in the 1563 nm resonance. Inset: Probe spectral measurements of a resonant mode at 0 mW and 4.25 mW on-chip pump powers.}
	\label{PumpShift}
\end{figure}

\begin{figure}
	\centering
	\includegraphics[width = \linewidth]{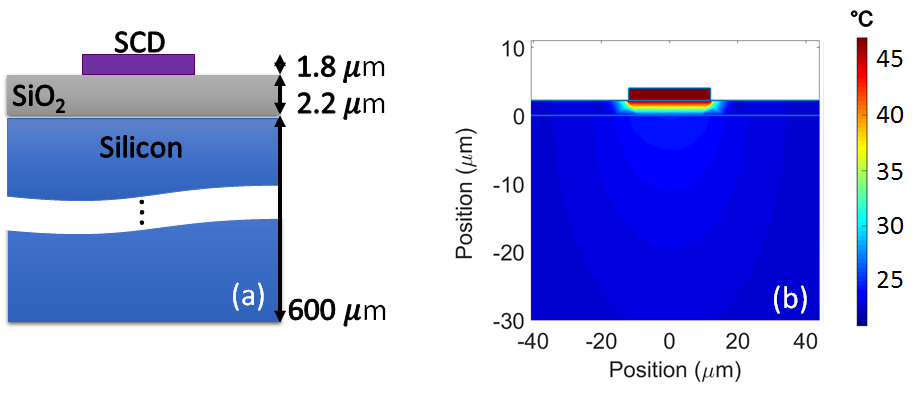}
	\caption{Simulation of thermal diffusion in the hybrid diamond-on-silica-on-silicon stack showing high confinement in the printed diamond micro-disk. a) Schematic of the material stack, b) thermal simulation close to the micro-disk region.}
	\label{Thermal}
\end{figure}

The propagation losses of the optical modes supported in the diamond micro-disk comprise of scattering and absorption components.  Since scattering is dominated by resonator sidewall roughness, it is expected this should be strongly correlated with absorption losses, dominated by surface state absorption.  Therefore a resonant mode with high round-trip propagation losses was selected to optically pump the device to maximise absorption and hence thermo-optic tuning of the cavity.  The micro-disk can be addressed in a pump/probe setup to decouple the optical signal required for tuning the resonance position and the probe beam used to measure the effective device transmission spectrum.  A resonance at a wavelength of 1563 nm, was selected for the pump, and simultaneous measurement of the effective transmission spectrum was taken using the probe beam. The effective resonant shift of the probe measurement is given as a function of the on-chip pump power in the tuning resonance in Figure \ref{PumpShift}.  
Resonant wavelength tuning up to 450 pm was demonstrated for the maximum available pump power of 4.25 mW on-chip. The inset of Figure \ref{PumpShift} shows the probe laser measurement of a single peak that has been shifted. Given a thermo-optic coefficient of 1.5 $\times$10$^{-5}$ for SCD, this translates to a uniform internal temperature of the micro-disk of 46$^o$C.  Figure \ref{Thermal} shows a finite element model of the diamond-on-silica-on-silicon cross section at the maximum measured temperature of 46$^o$C, assuming the background environment to be at room temperature.  The high thermal resistance of the silica layer that the diamond micro-disk is printed onto allows good confinement of the thermal energy to the diamond material, supporting the thermo-optic tuning effects observed in the measurements.

\section{Conclusion}
In conclusion, micro-fabrication and transfer printing techniques have been developed that enable the heterogeneous integration of monolithic diamond optical devices with non-native substrates. A diamond micro-disk resonator was printed onto a silicon waveguide chip with high alignment precision and exhibiting loaded quality factors on the order of 3.1$\times$10$^{4}$, with a maximum value for one resonance at 1.05$\times$10$^{5}$. Separable thermal tuning and spectral measurement of the resonances was demonstrated, with \textasciitilde 450 pm shifts shown at an on-chip pump level of 4.25 mW. The thermal insulation inherent to the bonding onto silica limits thermal cross talk between devices with spacings in the order of tens of microns.  Electrically controlled thermal tuning devices are commonly employed in integrated optics \cite{Heater_2013} and would be straightfoward to implement for hybrid diamond on PIC devices.  Such localised electronic thermal tuning elements could be used for active and stable tuning of individual diamond devices to align resonances across several devices on a single chip.

\section{Funding Information}
This work was supported by the EPSRC [EP/P013597/1, EP/P013570/1, EP/L015315/1, EP/L021129/1] and Fraunhofer UK. The authors acknowledge the efforts of the staff of the James Watt Nanofabrication Centre at the University of Glasgow. 


\end{document}